          [2001/04/25 1.1 (PWD)]
\documentclass[a4paper,twocolumn]{esapub2005} 
\usepackage{times}
\usepackage{natbib}
\usepackage{graphicx}

\title{MGGPOD: A Monte Carlo Suite for Gamma Ray Astronomy -- Version 1.1}
\author{G. Weidenspointner}
\affil{Centre d'Etude Spatiale des Rayonnements, BP 4346, 31028
Toulouse Cedex 4, France}
\author{S.J. Sturner}
\affil{NASA Goddard Space Flight Center, Code 661, Greenbelt, MD
20771, USA}
\author{E.I. Novikova}
\affil{Navy Research Laboratory, Washington, DC, USA}
\author{M.J. Harris}
\affil{CSNSM, IN2P3, Orsay Paris, France}
\author{A. Zoglauer}
\author{C.B. Wunderer}
\affil{Space Sciences Laboratory, University of California Berkeley,
CA 94720, USA}
\author{R.M. Kippen}
\affil{Los Alamos National Laboratory, NM, USA}
\author{P. Bloser}
\affil{University of New Hampshire, Durham, NH, USA}
\author{Ch. Zeitnitz}
\affil{Universit\"at Mainz, Mainz, Germany}

\begin{document}


\maketitle

\begin{abstract}

%
We announce the forthcoming public release of Version 1.1 of MGGPOD, a
user-friendly suite of Monte Carlo codes built around the widely used
GEANT (Version 3.21) package. MGGPOD is capable of simulating {\sl ab
initio} the physical processes relevant for the production of
instrumental backgrounds. These processes include the build-up and
delayed decay of radioactive isotopes as well as the prompt
de-excitation of excited nuclei, both of which give rise to a plethora
of instrumental gamma-ray background lines in addition to continuum
backgrounds. A detailed qualitative and quantitative
understanding of instrumental backgrounds is
crucial for most stages of high-energy astronomy missions.

Improvements implemented in Version 1.1 of the proven
MGGPOD Monte Carlo suite include: additional beam geometry options,
the capability of modelling polarized photons, additional output formats
suitable e.g.\ for event reconstruction algorithms, improved neutron
interaction cross sections, and improved treatment of the radioactive
decay of isomeric nuclear states.

The MGGPOD package and documentation are publicly available for
download from {\sl http://sigma-2.cesr.fr/spi/MGGPOD/}.
\end{abstract}


\section{Introduction}
\label{intro}

Intense and complex instrumental backgrounds, against which the much
smaller signals from celestial sources must be discerned, are a
notorious problem for low energy gamma-ray and hard X-ray astronomy.
Therefore a detailed qualitative and
quantitative understanding of instrumental line and continuum
backgrounds is crucial for most stages of gamma-ray astronomy
missions, 
ranging from the
design, development, and performance prediction phases through calibration
and response generation to data reduction. A
promising approach for obtaining quantitative estimates of instrumental
backgrounds is {\sl ab initio} Monte Carlo simulation \citep[see
e.g.][]{Dean03}. 

We have developed a suite of Monte Carlo packages, named MGGPOD
\citep{weidenspointner_mggpod}, that supports this type of simulation. The
original MGGPOD Monte Carlo suite (Version 1.0) and documentation have
been publicly available for download from {\sl
http://sigma-2.cesr.fr/spi/MGGPOD/} since 2004. First applications
include the successful modelling of the instrumental backgrounds of
the TGRS Ge spectrometer onboard Wind \citep{weidenspointner_mggpod},
the SPI Ge spectrometer onboard the INTEGRAL observatory
\citep{weidenspointner_mggpod_aa, weidenspointner_mggpod_seeon}, and
of the Reuven Ramaty High-Energy Solar Spectroscopic Imager
\citep{wunderer_rhessi}. 

Since then MGGPOD has found an increasing number of applications
including studies of the various instrument concepts for an Advanced
Compton Telescope (ACT) \citep[see
e.g.][]{Boggs06a,Boggs06b,wunderer_act}, for focal plane detectors of
Laue lens gamma-ray telescopes such as MAX or the Gamma-Ray Imager
(GRI) \citep{weidenspointner_bonifacio, wunderer_bonifacio,
Knoedlseder_GRI}, and for coded mask hard X-ray telescopes such as EXIST
\citep{Garson05,Garson06}. These studies required additions to and
improvements of the original MGGPOD Version 1.0, which will be made
available to the general community in the forthcoming public release
of Version 1.1 of MGGPOD after completing rigorous testing by the ACT
and GRI collaborations. 

In this paper we provide an overview of the capabilities, usage, and
structure of the MGGPOD package, as well as a description of the
additions and improvements implemented in Version 1.1.


\section{The MGGPOD Monte Carlo Simulation Suite}
\label{mggpod}

High-energy astronomy instruments are operated in an intense and
complex radiation environment \citep[see
e.g.][]{weidenspointner_mggpod}. The MGGPOD Monte Carlo suite allows
{\sl ab initio} simulations of instrumental backgrounds -- including
the many gamma-ray lines -- arising from interactions of the various
ambient particle and photon radiation fields within the instrument
and spacecraft materials. It is possible to simulate both prompt
instrumental backgrounds, such as energy losses of cosmic-ray
particles and their secondaries, as well as delayed instrumental
backgrounds, which are due to the decay of radioactive isotopes
produced in nuclear interactions. Of course, MGGPOD can also be used
to study the response of gamma-ray instruments to astrophysical and
calibration sources. The MGGPOD suite is therefore an ideal Monte
Carlo tool for high-energy astronomy.
A detailed description of the physics implemented in the original
MGGPOD Monte Carlo suite (Version 1.0) can be found in
\citet{weidenspointner_mggpod}. The documentation available from the
MGGPOD web site provides comprehensive practical advice for users.

\subsection{Capabilities and Functionalities}
\label{cap_func}


MGGPOD is a suite of five closely integrated Monte Carlo packages,
namely {\bf MG}EANT, {\bf G}CALOR, {\bf P}ROMPT, {\bf O}RIHET, and
{\bf D}ECAY. The MGGPOD suite resulted from a combination of the
NASA/GSFC MGEANT \citep{Sturner03} and the University of Southampton
GGOD
\citep{Dean03} Monte Carlo codes, which 
were supplemented with the newly developed PROMPT package. All these
packages are based on the widely used GEANT Detector Description and
Simulation Tool (Version 3.21) created and supported at
CERN\footnote{see {\sl http://wwwinfo.cern.ch/asd/geant/}},
which is designed to simulate the passage of
elementary particles through an experimental setup.

In the following, we provide a synopsis of the capabilities and
functions of the five packages that constitute the MGGPOD suite:
\begin{itemize}
\item MGEANT \citep{Sturner03} is a multi-purpose simulation package
that was created to increase the versatility of the GEANT simulation
tool. A modular, ``object oriented'' approach was pursued, allowing
for rapid prototyping of detector systems and easy generation of most
of the radiation fields relevant to high-energy astronomy. Within the
MGGPOD suite, MGEANT (i.e.\ GEANT) stores and transports all
particles, and treats electromagnetic interactions from about 1~keV to
a few TeV\footnote{By default, the low-energy cutoff in GEANT is
10~keV. This cutoff energy can, however, be lowered, as e.g.\
described in the MGGPOD documentation. We have successfully lowered
the cutoff energy down to 1~keV.}. MGEANT provides the option to use
the GLECS and GLEPS packages to take into account the energy of bound
electrons and photon polarization in Rayleigh and Compton scatterings
(see Sec.~\ref{mggpod1.1}).
The MGEANT simulation package and a user manual
are available at a NASA/GSFC web site\footnote{see {\sl
http://lheawww.gsfc.nasa.gov/docs/gamcosray/legr/mgeant/ mgeant.html}}.
\item GCALOR 
\citep{Zeitnitz_Gabriel94} simulates
hadronic interactions down to 1~MeV for nucleons and charged pions and
down to thermal energies ($10^{-5}$~eV) for neutrons. Equally
important, this package\footnote{see {\sl
http://www.staff.uni-mainz.de/zeitnitz/Gcalor/gcalor.html}}
provides access to the energy deposits from all hadronic interactions as well
as to isotope production anywhere in the simulated setup.
\item PROMPT simulates prompt photon emission associated with the
de-excitation of excited nuclei produced by neutron capture, inelastic
neutron scattering, and spallation.
\item ORIHET, originally developed for the GGOD suite \citep{Dean03} and
improved for MGGPOD, calculates the build-up and decay of activity in
any system for which the nuclide production rates are known. Hence
ORIHET can be used to convert nuclide production rates, determined
from simulations of cosmic-ray irradiation, to decay rates. These are
required input for simulating the radioactive decays giving rise to
delayed background.
\item DECAY, again originally developed for GGOD and improved for our
purposes, enables MGGPOD to simulate radioactive decays.
\end{itemize}

\subsection{Structure}
\label{structure}

\begin{figure}
\centering
\includegraphics[width=5.25cm,angle=270.,bbllx=125pt,bblly=127pt,bburx=485pt,bbury=665pt,clip=]{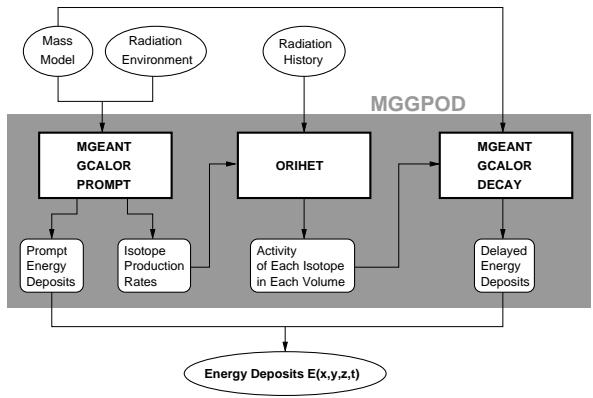}
\caption{A flow chart illustrating the overall
structure of the MGGPOD Monte Carlo simulation suite. The various
simulation packages (shown in boxes) and input and output files (shown
in ellipses and round-edged boxes) are explained in the
text\label{mggpod_flow_chart}}
\end{figure}

The overall structure of the MGGPOD package is illustrated in
Fig.~\ref{mggpod_flow_chart}. Depending on the simulated radiation
field or high-energy photon source distribution one or three steps,
requiring two or three input files, are needed to obtain the resulting
energy deposits in the detector system under study. In general, it is
advisable to simulate each component of the radiation environment
separately.
MGGPOD distinguishes two classes of radiation fields. 
\begin{itemize}
\item Class~I comprises radiation fields for which only prompt energy
deposits are of interest, such as celestial or laboratory gamma-ray
sources or cosmic-ray electrons. 
\item Class~II comprises radiation fields for which in addition
delayed energy deposits resulting from the activation of radioactive
isotopes need to be considered. Examples for Class~II fields are
cosmic-ray protons, geomagnetically trapped protons, or neutrons
leaking from the Earth's atmosphere.
\end{itemize}

For both of these classes, the simulation of the prompt energy
deposits requires two inputs: a mass model, and a model of the
simulated radiation field. The mass model is a detailed computer
description of the experimental setup under study. It specifies the
geometrical structure of instrument and spacecraft, the atomic and/or
isotopic composition of materials, and sets parameters that influence
the transport of particles in different materials. Each component of
the radiation environment (and analogously for gamma-ray sources) to
which the instrument is exposed is characterized by three quantities:
the type of the incident particles, and their spectral and angular
distributions. The prompt energy deposits are written to an output
event file; in case of a Class~II radiation field there is an
additionial output file in which all the nuclei produced in hadronic
interactions are recorded.

To simulate delayed energy deposits (Class~II radiation field only) two
additonal steps need to be taken. These require as input the time
history of the radiation field which is
responsible for the activation, and the isotope production rates
obtained previously during the simulation of prompt background.
Based on this information, first the activity of each isotope produced in
each structural element of the mass model 
is determined. Then these activities are used to simulate the delayed
energy deposits in the instrument due to radioactive decays in each
volume of the mass model.

Combining prompt and delayed energy
deposits from each component of the radiation environment and
gamma-ray sources, it is possible to obtain the energy deposited
in the system as a function of position and time. In particular, it is
e.g.\ possible to obtain background energy spectra and rates for each
detector element, and to assess the individual contributions of the
various background components and mass model volumes, which is highly
valuable when designing new instruments.



\section{Additions and Improvements for MGGPOD Version 1.1}
\label{mggpod1.1}

The additions to and improvements of the original MGGPOD Version 1.0 were
driven by the requirements for studying various instrument concepts
for an Advanced Compton Telescope (ACT), Laue lens gamma-ray telescope
focal plane detectors, and coded mask hard \mbox{X-ray} telescopes.
The original version of MGGPOD was mainly tested by modelling
gamma-ray instruments that employ monolithic Ge detectors, such as the
TGRS Ge spectrometer onboard Wind \citep{weidenspointner_mggpod}, the
SPI Ge spectrometer onboard the INTEGRAL observatory
\citep{weidenspointner_mggpod_aa, weidenspointner_mggpod_seeon}, and
of the Reuven Ramaty High-Energy Solar Spectroscopic Imager
\citep{wunderer_rhessi}. Future high-energy instruments will use
highly segmented or pixellated detectors. In particular detectors for
Compton telescopes aim not only at recording the energy of each
incoming photon, but rather at recording the complete interaction
sequence. Furthermore, a large number of different detector materials
is being considered for future instrumentation (e.g.\ Si, Xe, CZT, CdTe);
hadronic interactions and nuclear de-excitations for these were not
well treated in Version 1.0 of MGGPOD.

All recent upgrades of the original MGGPOD suite will soon be publicly
released in Version 1.1, after completing rigorous testing by the ACT
and GRI collaborations. The additions and improvements for MGGPOD
Version 1.1 are as follows.
\begin{itemize}
\item We introduced new beam geometries. It is now possible to model
radiation fields whose intensity varies with zenith angle -- a
requirement for simulating e.g.\ Earth leakage radiations in low-Earth
orbit (gamma-rays, neutrons, \dots), or atmospheric background at
balloon altitudes. Models can e.g.\ be generated with the ACTtools
software available at {\sl
http://public.lanl.gov/mkippen/actsim/glecs/}.
Also, new beam options were introduced to enable the simulation of
photon beams from Laue diffraction lenses or hard X-ray mirrors.
\item 
Additional output formats optimized for use with event reconstruction
algorithms have been implemented. Among other information, 
it is possible to record the complete interaction information for all
involved particles in passive as well as active (i.e.\ detector or
veto shield) instrument materials. This is necessary e.g.\ for the
creation of (probability based) response files for event
reconstruction algorithms for Compton telescopes \citep[e.g.\ MEGAlib,
see][]{Zoglauer06}. It is also possible to record vertex (i.e.\
initial position) and momentum vector for each simulated particle. This
capability to retrace the origin of each event has also been helpful
for optimizing e.g.\ the design of active and passive shielding for
high-energy instrumentation.
\item One of the many exciting science goals of the
next generation of high-energy instruments is to measure
polarization. To include the effects of polarized incident photons in
the Compton and Rayleigh scattering processes, the GLEPS
package\footnote{see {\sl
http://public.lanl.gov/mkippen/actsim/glecs/}} is now available in
MGGPOD. GLEPS \citep{McConnell_Kippen04}
is an extension of the GLECS GEANT3 Low-Energy Compton Scattering
Package \citep{Kippen04},
which takes into account the energy of bound electrons in photon
scattering processes. 
\item The standard set of neutron cross sections
available for GCALOR does not cover all elements/isotopes relevant for
studying gamma-ray instrumentation. For selected elements/isotopes, we
converted into GCALOR format the respective
evaluated ENDF/B and JENDL neutron cross sections (complemented by
other databases for isotopes of Zn missing in ENDF/B and JENDL).  A
slight modification of GCALOR was necessary to
make sure that all these new cross section files can be read in
properly.
\item For selected elements/isotopes relevant for high-energy
instrumentation (including Ge, Si, Xe, Cd, Zn, Te, Cs, I) we generated
the PROMPT data files required to model de-excitations after neutron
capture and inelastic neutron scattering. (De-excitations after
spallations have already been modelled in Version 1.0 based on
statistical considerations as described in
\citep{weidenspointner_mggpod}.)
\item In Version 1.0 of MGGPOD it was assumed that
nuclei in isomeric states decay exclusively through internal
transition. However, this is not always true -- some isomeric states can
also beta-decay. The beta decay channel of isomeric states has now
been implemented, improving the simulation of delayed background due to
radioactive decays in common detector materials such as CZT.
\item Last but not least, we introduced many checks to avoid improper
input by the user or in data files (e.g.\ DECAY or PROMPT), which if
not recognized may
lead to run-time errors that are hard to diagnose. In case improper
input is identified, error or warning messages are written to the log
file, and the run is terminated if necessary.
\end{itemize}


\section{Summary}

The MGGPOD Monte Carlo suite is an ideal tool for supporting the
various stages of high-energy astronomy missions, ranging from the
design, development, and performance prediction through calibration
and response generation to data reduction. Version 1.0 of the MGGPOD
software and documentation are publicly available for download at
CESR. The package has been, and is being, successfully applied to an
increasing number of past and present gamma-ray and hard X-ray
missions. New applications often entail requirements for new
functionalities, hence the MGGPOD suite is evolving continuously.  The
additions and improvements required for studying various instrument
concepts for an Advanced Compton Telescope, Laue lens gamma-ray
telescope focal plane detectors, and coded mask hard X-ray telescopes
will be made available to the community in the forthcoming release of
Version 1.1 of MGGPOD.

\end{document}